\documentclass[aps,preprint,showpacs,nofootinbib]{revtex4}
\usepackage{amsfonts}
\usepackage{amsmath}
\usepackage{amssymb}
\usepackage{relsize}
\usepackage{graphicx}
\usepackage{pdfpages}
\usepackage{subfigure}
\usepackage{float}
\usepackage[caption = false]{subfig}
\usepackage{color}
\usepackage{multirow}
\begin{document}

\title{Averaged electron densities of the helium-like atoms}

\author{Evgeny Z. Liverts}
\affiliation{Racah Institute of Physics, The Hebrew University, Jerusalem 91904, Israel}


\author{Rajmund Krivec}
\affiliation{Department of Theoretical Physics, J. Stefan Institute, 1000 Ljubljana, Slovenia}

\begin{abstract}
Different kinds of averaging of the wavefunctions/densities of the two-electron atomic systems are investigated.
Using the Pekeris-like method \cite{LEZ1}, the ground state wave functions $\Psi$
of the helium-like atoms with nucleus charge $1\leq Z\leq5$ are calculated in a few coordinate systems
including the hyperspherical coordinates $\left\{R,\alpha,\theta\right\}$.
The wave functions $\Psi_{av}(R)$ of the hyperspherical radius $R$ are calculated numerically by averaging $\Psi$ over the hyperspherical angles $\alpha$ and $\theta$.
The exact analytic representations for the relative derivatives $\Psi_{av}'(0)/\Psi_{av}(0)$ and $\Psi_{av}''(0)/\Psi_{av}(0)$ are derived.
Analytic approximations very close to the actual $\Psi_{av}(R)$ are obtained.

Using actual wave functions $\Psi$, the one-electron densities $\rho(r)$ are calculated as functions of the electron-nucleus distance $r$. The relevant derivatives $\rho'(0)/\rho(0)$ and $\rho''(0)/\rho(0)$ characterizing the behavior of $\rho(r)$ near the nucleus are calculated numerically. Very accurate analytical approximations, representing the actual one-electron density both near the nucleus and far away from it, are derived.
All the analytical and numerical results are supplemented with tables and graphs.
\end{abstract}


\maketitle

\section{Introduction}\label{S0}

The only exactly solvable problems in atomic and molecular quantum mechanics are the one-electron problems of $H$, and within the Born-Oppenheimer approximation, $H_2^+$. To get deeper insight into the analytical properties of the one-electron density $\rho(r)>0$ of many-electron systems \cite{FOUR0}, it is desirable to investigate those features of this density that carry over to the many-electron case. Using the Wolfram Mathematica codes \cite{LEZ1} we investigate the properties of the one-electron densities/wavefunctions of the two-electron atom/ions. The relevant behavior at the nucleus described by the relative derivatives
 $\rho'(0)/\rho(0)$ and $\rho''(0)/\rho(0)$
, as well as the asymptotic behavior when the electron-nucleus distance $r$ approaches infinity, are studied both numerically and analytically.
Besides of the one-electron density $\rho(r)$ representing the form of the averaged many-electron system, we also examine other kinds of averaging.

We shall consider the two-electron atomic systems which can serve as excellent models for testing and verifying quantum theories.
It is well-known that the wave function (WF) of many-electron (and the two-electron, in particular) system represents a function of many variables.
In order to proceed to the simply interpretable case of a single variable, several methods can be applied.

The first method is to set specific values of the angular variables in the hyperspherical coordinates.
In particular, the S-state WF of the helium-like atoms $\Psi$ reduces to a function of three variables, e.g., the electron-nucleus distances $r_1,r_2$ and the electron-electron distance $r_{12}$ or  the equivalent hyperspherical coordinates
\begin{equation}\label{I1}
R=\sqrt{r_1^2+r_2^2},~~~~\alpha=2\arctan \left(r_2/r_1\right),~~~~\theta=\arccos\left[(r_1^2+r_2^2-r_{12}^2)/(2r_1r_2)\right].~~~
\end{equation}
Setting the hyperspherical angles $\alpha=0, \theta=\pi/2$ or $\alpha=\pi/2, \theta=0$, we obtain the configurations 
of the electron-nucleus or the electron-electron coalescence, respectively.
Recent studies of these states can be found in Ref.\cite{LEZ2} (see also references therein).

The other methods are connected with some kind of averaging over specific sets of variables.
Representing the WF of the $n$-electron atomic system in the hyperspherical coordinates and integrating subsequently over the angular coordinates, we obtain a function of the hyper-radius $R=\sqrt{\sum_{i=1}^n{r_i^2}}$ only, where $r_i$ is the distance between the nucleus and the $i$-th electron (see, e.g., \cite{HOF1} and references therein).

These methods of averaging can be described as follows (see, e.g., \cite{HOF2}). The first step consists in introducing the so called diagonal of the spinless one-electron density matrix
\begin{equation}\label{I2}
\rho(\textbf{r}_1,\textbf{r}_1')=\int \Psi^*(\textbf{r}_1',\textbf{r}_2,...,\textbf{r}_n)\Psi(\textbf{r}_1,\textbf{r}_2,...,\textbf{r}_n)
d^3\textbf{r}_2...d^3\textbf{r}_n ,
\end{equation}
to within a constant multiplicative factor. Note that the radius-vector $\textbf{r}_1$ can represent any of the $n$ electrons of the considered atomic system.
Then the averaged function, called the one-electron WF, is introduced as $\psi_{av}(r)=\sqrt{\rho_{av}(r)}$,
where the one-electron density
\begin{equation}\label{I4}
\rho_{av}(r)=\int_0^\pi d \theta'\int_0^{2\pi} \rho(r,\theta',\phi')\sin \theta'd\phi'
\end{equation}
is the spherically averaged electron density $\rho_(r,\theta',\phi')\equiv\rho(\textbf{r},\textbf{r})$.
We have denoted the azimuthal angle of the spherical polar coordinates of the electron by $\theta'$ in order not to confuse it with hyperspherical angle $\theta$ which is the angle between the radius-vectors $\textbf{r}_1$ and $\textbf{r}_2$ of the electrons of the helium-like atoms.
We would like to highlight the fact that here the averaged WF is a function of the electron-nucleus distance $r$, unlike the averaged WF mentioned before, which was a function of the hyperspherical radius $R$.

In the subsequent sections we shall study the principal properties of both kinds of the averaged electron density/wavefunction describing the two-electron atomic systems in S-states.

\section{Averaging over the hyperspherical angles}\label{S1}

The authors of the paper \cite{HOF1} considered the $n$-electron atomic system. They have derived the general expression for the relative first derivative $\psi'(0)/\psi(0)$ at the nucleus obtained by averaging the $n$-electron WF over the hyperspherical angles.
In this section we shall derive the analytic expression for the first, as well as for the second relative derivative $\psi''(0)/\psi(0)$ at the nucleus of the two-electron atomic systems in S-state. The analytic representation for the averaged WF $\psi(R)$, very close to the actual one, will be derived and analyzed too.

The Schr\"{o}dinger equation for the helium-like atom/ions with energy $E$ and infinitely massive nucleus of charge $Z$ reads
\begin{equation}\label{K1}
\left(-\frac{1}{2}\Delta+V\right)\Psi=E~\Psi.
\end{equation}
The Laplacian in the hyperspherical coordinates is of the form
\begin{equation}\label{K2}
\Delta=\frac{\partial^2}{\partial R^2}+\frac{5}{R}\frac{\partial}{\partial R}-\frac{1}{R^2}\Lambda^2,
\end{equation}
where $\Lambda^2$ is the hyperspherical angular momentum operator projected on S states.
The inter-particle Coulomb interaction
\begin{equation}\label{K3}
V=-\frac{Z}{r_1}-\frac{Z}{r_2}+\frac{1}{r_{12}}
\end{equation}
can be expressed in the hyperspherical coordinates using the transform
which is inverse to (\ref{I1}).
Let's denote the average of the function $f(R,\alpha,\theta)$ over the hyperspherical angles $\alpha,\theta$ as
\begin{equation}\label{K4}
f_{av}(R)=\int_0^{\pi}d \theta\int_0^\pi f(R,\alpha,\theta)\sin^2\alpha \sin \theta~ d \alpha .
\end{equation}
Then, averaging the Schr\"{o}dinger equation (\ref{K1}) reduces it to the form
\begin{equation}\label{K5}
\left(\frac{\partial^2\Psi}{\partial R^2}\right)_{av}+\frac{5}{R}\left(\frac{\partial\Psi}{\partial R}\right)_{av}-2\left(V \Psi\right)_{av}=-2 E \Psi_{av}.
\end{equation}
When deriving Eq.(\ref{K5}) we took into account the non-trivial equation
\begin{equation}\label{K6}
\left(\Lambda^2 \Psi\right)_{av}=0 ,
\end{equation}
which is valid for all $R$ \cite{HOF1}.
To derive the properties of $\Psi_{av}(R)$ at the triple coalescence point $R=0$, it is natural to use the Fock expansion \cite{FOCK} (see also \cite{LEZ2})
\begin{equation}\label{K7}
\Psi({R},\alpha,\theta)=\sum_{k=0}^{\infty}{R}^k\sum_{p=0}^{[k/2]}\psi_{k,p}(\alpha,\theta)(\ln {R})^p .
\end{equation}
Substituting this expansion into the averaged Schr\"{o}dinger equation (\ref{K5}) and equating coefficients for the same powers of $R$ and $\ln R$, we obtain the recurrence relations (RER)
\begin{equation}\label{K8}
k(k+4)(\psi_{k,p})_{av}+2(k+2)(p+1)(\psi_{k,p+1})_{av}+(p+1)(p+2)(\psi_{k,p+2})_{av}-2(\mathcal{V} \psi_{k-1,p})_{av}+2E(\psi_{k-2,p})_{av}=0
\end{equation}
for the averaged angular coefficients $(\psi_{k,p})_{av}$. Here $\mathcal{V}\equiv\mathcal{V}(\alpha,\theta)= R~V$, where $V$ is the potential (\ref{K3}). The explicit form of $\mathcal{V}$ can be found in Appendix \ref{SA}. Using the definition (\ref{K4}), the Fock expansion (\ref{K7}) and the RER (\ref{K8}), we obtain:
\begin{equation}\label{K9}
\Psi_{av}(0)=(\psi_{0,0})_{av}=\pi,~~~~~\Psi_{av}'(0)=(\psi_{1,0})_{av},~~~~~\Psi_{av}''(0)=2(\psi_{2,0})_{av}.~~~~~~
\end{equation}
We took into account the fact that all of the angular coefficients (AC) calculated so far were derived under the condition $\psi_{0,0}=1$.
From here and in what follows, the notation $f'(x),f''(x),...$  refers to the first, second and so on derivatives of $f(x)$ with respect to $x$.
The explicit forms of the angular coefficients $\psi_{k,p}$ can be found in \cite{LEZ2}. This yields:
\begin{equation}\label{K10}
\varrho\equiv\frac{\Psi_{av}'(0)}{\Psi_{av}(0)}=-\frac{16\left(4Z-\sqrt{2}\right)}{15\pi},
\end{equation}
\begin{equation}\label{K11}
\lambda\equiv\frac{\Psi_{av}''(0)}{\Psi_{av}(0)}=\frac{1-2E}{6}-\frac{2(\pi+2)}{3\pi}Z+\frac{2(\pi+4)}{3\pi}Z^2.
\end{equation}
The details of deriving the results (\ref{K10}) and (\ref{K11}) can be found in  Appendix \ref{SA}. Expression (\ref{K10}) for the first relative derivative (logarithmic derivative) is certainly coincident with that obtained in Ref.\cite{HOF1}. We have denoted the first and second relative derivatives in Eqs.(\ref{K10}) and (\ref{K11}) by $\varrho$ and $\lambda$, respectively, given the necessity of utilizing them in what follows.

It is worth noting that our calculations reveal the vanishing averaged AC $\psi_{2,1}(\alpha,\theta)$, that is $(\psi_{2,1})_{av}=0$. This property prevents the divergence of $\Psi_{av}''(0)$.
On the other hand, using the already calculated AC (see \cite{LEZ2}), it can be shown that $(\psi_{3,1})_{av}\neq0$ (as well as $(\psi_{4,1})_{av}\neq0$). This tells us that there is no guarantee that $\Psi_{av}'''(0)$ exists and is finite.

Using the Pekeris-like method and codes \cite{LEZ1} we can calculate the "actual" WFs $(\Psi)$ of the helium-like atom/ions in the S-state. This enables us to calculate numerically the averaged WF, $\Psi_{av}$, according to definition (\ref{K4}). Using such a calculation it is easy to verify the following features of averaging  (\ref{K4}):
\begin{equation*}
\frac{d\Psi_{av}(R)}{d R}=\left(\frac{\partial \Psi}{\partial R}\right)_{av},~~~~~~~
\frac{d^2\Psi_{av}(R)}{d R^2}=\left(\frac{\partial^2 \Psi}{\partial R^2}\right)_{av}.~~~
\end{equation*}
Thus, the averaged Schr\"{o}dinger equation (\ref{K5}) can be rewritten in the form
\begin{equation}\label{K12}
\Psi_{av}''+\frac{5}{R}\Psi_{av}'-2\left(V \Psi\right)_{av}=-2 E \Psi_{av}.
\end{equation}
Our aim is to cast Eq.(\ref{K12}) in the form of an analytically solvable differential equation.
To this end we propose the following approximation
\begin{equation*}
(V \Psi)_{av}=\left(\frac{b_1}{R}+b_0 \right)\Psi_{av}'(R)+\left(\frac{a_1}{R}+a_0\right)\Psi(R),
\end{equation*}
where $a_0,a_1,b_0$ and $b_1$ are some parameters.
Our calculations with the use of "actual" WFs show not only the existence but also high efficiency of such an approximation.
Substitution of this approximation into Eq.(\ref{K12}) leads to the general differential equation
\begin{equation}\label{K13}
\Psi_{av}''+\left(\frac{B_1}{R}+B_0\right)\Psi_{av}'+\left(\frac{A_1}{R}+A_0\right)\Psi_{av}=0,
\end{equation}
where $A_0,A_1,B_0$ and $B_1$ are as yet undetermined parameters.
The general solution of Eq.(\ref{K13}) is of the form
\begin{equation}\label{K14}
\Psi_{av}(R)=\exp\left[-\frac{1}{2}R\left(B_0+\zeta\right)\right]\left[c_1 U\left(\eta,B_1,\zeta R\right)+c_2 L_{-\eta}^{B_1-1}(\zeta R)\right],
\end{equation}
where $U(a,b,z)$ is the confluent hypergeometric function of the second kind (or Tricomi function), and $L_{-a}^{b-1}(z)$ is the generalized Laguerre function. In the representation (\ref{K14}) we have introduced the following notation:
\begin{equation}\label{K15}
\zeta=\sqrt{B_0^2-4 A_0},~~~~~~~~~~\eta=\frac{B_1}{2}+\frac{B_0 B_1-2A_1}{2\zeta}.
\end{equation}
It is easy to verify that the behavior of the Tricomi function $U\left(\eta,B_1,\zeta R\right)$ near the triple coalescence point $R=0$ does not allow us to apply this special function to construct the averaged WF with correct behavior in the vicinity of nucleus ($\Psi_{av}(0)=const$).
Whence, putting $c_1=0$ in the general solution (\ref{K14}), we obtain the physical solution
\begin{equation}\label{K16}
\Psi_{av}(R)=\exp\left[-\frac{1}{2}R\left(B_0+\zeta\right)\right]L_{-\eta}^{B_1-1}(\zeta R),
\end{equation}
to within a constant multiplicative factor.
The power series expansion of the function (\ref{K16}) together with the exact values of the relative derivatives $\varrho$ and $\lambda$, defined by Eqs.(\ref{K10}) and (\ref{K11}),  enable us to obtain the following two coupling equations for the four parameters $A_0,A_1,B_0,B_1$ appearing in the approximation WF (\ref{K16}):
 \begin{equation}\label{K17}
A_1=-B_1\varrho,~~~~~~~~~~~~A_1(A_1+B_0)-B_1 A_0=\lambda~B_1(B_1+1).
\end{equation}
The two remaining parameters (together with the first two, of course) can be calculated by fitting the averaged "actual"  WF \cite{LEZ1} in the specific range $[0,R_m]$ of the substantial decrease of this function.
The sets of parameters calculated by the method described above are presented in Table \ref{T1}.
It is worth noting that the Laguerre function $L_{-\eta}^{B_1-1}(\zeta R)$ diverges like $\exp(\zeta R)$ as $R\rightarrow \infty$. However, the exponential factor $\exp[-R(B_0+\zeta)/2]$ in Eq.(\ref{K16}) defines the asymptotic behavior of $\Psi_{av}(R)$ as proportional to $\exp[-R(B_0-\zeta)/2]$.
It follows from definition $\zeta$ in Eq.(\ref{K15}) that the exponent $-R(B_0-\zeta)/2$ will be negative for $A_0>0$. It is seen from Table \ref{T1} that the latter condition is met for all of the presented atoms. This implies that the averaged approximation (\ref{K16}) is not divergent on the full interval $R\in [0,\infty]$,  not only on the fitting range $R\in [0,R_m]$.

\begin{figure}
\centering
\caption{The ground state of helium. \emph{\textbf{(a)}} The WF averaged over the hyperspherical angles.~\emph{\textbf{(b)}} The logarithmic estimate of the accuracy of the WF approximation in the form represented by Eq.(\ref{K16}).}
\includegraphics[width=8.0in]{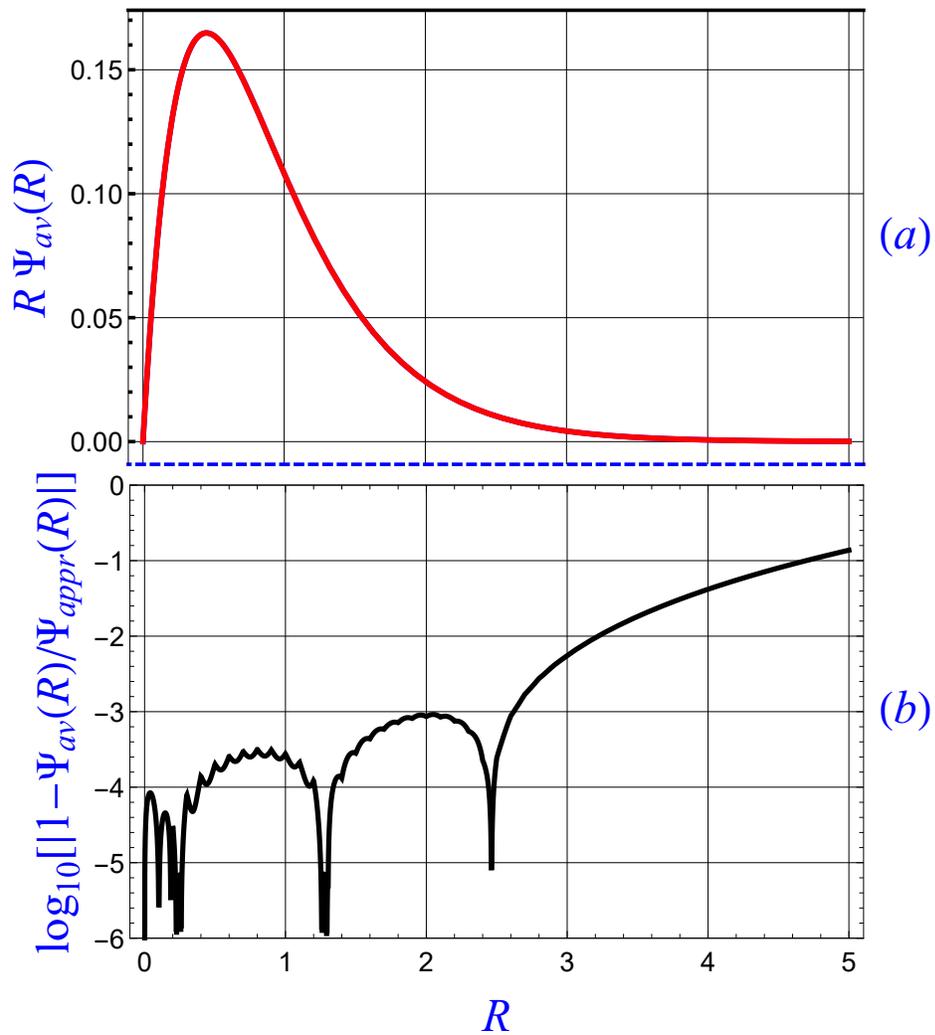}
\label{F1}
\end{figure}

In Fig.\ref{F1} we have presented functions $R\Psi_{av}(R)$ and $R \Psi_{appr}(R)$ for the ground state of helium. The function  $\Psi_{av}(R)$ was obtained by averaging the Pekeris-like wave function \cite{LEZ1} with $\Omega=25$ and  $\Psi_{av}(0)=1$, whereas function $ \Psi_{appr}(R)$  was calculated by analytic representation (\ref{K16}) with parameters presented in Table \ref{T1}.
Using these parameters it is easy to verify that parameter $\zeta$ introduced by Eq.(\ref{K15}) is complex for some $Z$. Accordingly, the approximate WF (\ref{K16}) can be a complex function. However, function $\Psi_{appr}(R)=\widetilde{\Psi}_{appr}(R)/\widetilde{\Psi}_{appr}(0)$, giving $\Psi_{appr}(0)=1$, will be real.

Note that on Fig.\ref{F1}\emph{a} there is no visual difference between two presented functions due to the fact that they are very close to each other. This also follows from Fig.\ref{F1}\emph{b} where the logarithmic characteristic $\log_{10}\left|1-\Psi_{av}(R)/\Psi_{appr}(R)\right|$ of the accuracy of the approximation WF was demonstrated.
The averaged WF of helium was presented as an example.
The similar graphs for other two-electron atoms
exhibit a minor increase of
the accuracy of the approximate WF as $Z$ increases.

\section{One-electron density} \label{S2}

In the Introduction we have described in short the method of calculation of the one-electron density/wavefunction for the $n$-electron atomic system.
It can be shown that such a calculation for the two-electron atoms can be carried out in the most simple and effective way by using the coordinate system $\left\{r_1,r_2,\omega\right\}$ with $\omega=\cos \theta$.
The angle $\theta$ has been defined in Eq.(\ref{I1}) is the angle between the radius-vectors $\textbf{r}_1$ and $\textbf{r}_2$ of the electrons, with the origin of the coordinate system located at the nucleus. The volume element is:
\begin{equation}\label{J1}
d V=8\pi^2 r_1 r_2 r_{12}d r_1 d r_2 d r_{12}=8\pi^2 r_1^2 r_2^2 d r_1 d r_2 d \omega,~~~~~~~~~~\omega\in[-1,1].
\end{equation}
The Laplacian in the coordinate system $\left\{r_1,r_2,\omega\right\}$ reads:
\begin{equation}\label{J2}
\Delta=r_1^{-2}\frac{\partial}{\partial r_1}r_1^2\frac{\partial}{\partial r_1}+
r_2^{-2}\frac{\partial}{\partial r_2}r_2^2\frac{\partial}{\partial r_2}+\left(\frac{1}{r_1^2}+\frac{1}{r_2^2}\right)
\left[\left(1-\omega^2\right)\frac{\partial^2}{\partial \omega^2}-2\omega \frac{\partial}{\partial \omega}\right].
\end{equation}
The one-electron density of the helium-like atom with the WF $\Phi(r_1,r_2,\omega)\equiv\Psi(R,\alpha,\theta)$ can be calculated as follows
\begin{equation}\label{J3}
\rho(r)=\int_{-1}^1 d \omega \int_0^\infty \left|\Phi(r,r_2,\omega)\right|^2 r_2^2 dr_2,
\end{equation}
to within a constant multiplicative factor.

\subsection{The asymptotic behavior} \label{S2a}

It was noted in Ref.\cite{FROL1} that as follows from Ref.\cite{HOF2} at large electron-nucleus distance $r$ the one-electron asymptotic $\psi_{as}(r)=\sqrt{\rho_{as}(r)}$ of the two-electron atomic system can be written in the form
\begin{equation}\label{J4}
\psi_{as}(r)=A r^{\frac{Z-1}{b}-1}\exp(- b r),
\end{equation}
where $b=\sqrt{-2E-Z^2}$, and $A$ is some numerical constant. This fact was used in earlier works on photodetachment of the $ H^-$ ion (see,
e.g.,\cite{FROL2} and references therein).
It is stated in the paper \cite{FROL1} that they have tested the overall accuracy of the general formula (\ref{J4}). It was found that this simple formula represents to very good accuracy ($\approx 10^{-4}-10^{-5} a.u.$) the asymptotic behavior ($r\rightarrow \infty $) of the actual two-electron wavefunctions for all considered ions.
Using the Mathematica codes based on the Pekeris-like method \cite{LEZ1}, we have provided similar tests and have obtained the same results.
For comparing the asymptotic formula (\ref{J4}) and the "actual" one-electron WFs we applied the logarithmic derivative $\psi'(r)/\psi(r)$ which is independent on the multiplicative constant $A$.

It is shown in Appendix \ref{SB} that at large enough $r$ the one-electron density $\rho(r)$ obeys the equation
\begin{equation}\label{J5}
\rho''(r)+\frac{4}{r}\rho'(r)+4\left[\frac{2(Z-1)}{r}+Z^2+2 E\right]\rho(r)=0.
\end{equation}
The general solution of this equation reads
\begin{equation}\label{J6}
\rho(r)=\exp(-2 b r)\left[c_1 U\left(2-\frac{2(Z-1)}{b},4,4 b r\right)+c_2 L_{-2+\frac{2(Z-1)}{b}}^3(4 b r)\right],
\end{equation}
where $U(a,4,z)$ is the Tricomi function, $L_{-a}^3(z)$ is the generalized Laguerre function, and the parameter $b$ is defined in (and just after) Eq.(\ref{J4}). It can be verified that the leading term of the asymptotic ($r\rightarrow\infty$) expansion of the Laguerre function in Eq.(\ref{J6}) is proportional to $\exp(4 b r)$ (for $Z>1$ only). Hence, we should put $c_2=0$ in the general solution (\ref{J6}) to obtain the one-electron WF (to within a constant multiplicative factor)
\begin{equation}\label{J7}
\psi(r)= \exp(- b r) \left[U\left(2-\frac{2(Z-1)}{b},4,4 b r\right)\right]^{1/2},~~~~~~~~(Z>1)
\end{equation}
exponentially vanishing at infinity. It is worth noting that the leading term of the asymptotic expansion of the one-electron WF (\ref{J7}) is coincident with the function (\ref{J4}), including the case of $Z=1$.

We shall determine the real behavior of the one-electron density $\rho(r)$ by the use of the "actual" WFs \cite{LEZ1} for numerical calculations of $\psi(r)=\sqrt{\rho(r)}$ according to Eq.(\ref{J3}). Let's call such calculations AWFC.
Using the AWFC, we have found that both the one-electron densities (\ref{J7}) and the asymptotic  WFs (\ref{J4}), taken with a slightly shifted argument, enable us to extend significantly the range of their applicability.
For comparison we use the logarithmic derivative $y(r)=\psi'(r)/\psi(r)$ which is independent of the constant multiplicative factor. As the characteristic of accuracy of the analytic approximation $y_{appr}(r)$ we apply the function
\begin{equation}\label{J8}
\mathfrak{L}_k(\tau_k)=\log_{10}\left|1-y_{appr}(r-\tau_k)/y_{act}(r)\right|,~~~~~~(k=1,2)
\end{equation}
where $y_{act}(r)$ is obtained by the AWFC.
Subscripts $k=1$ and $k=2$ correspond to the approximate WFs of the form (\ref{J4}) and (\ref{J7}), respectively, whereas $\tau_k$ represents the shift of the argument $r$ associated with the corresponding approximate WF.

\begin{figure}
\centering
\caption{The ground state of helium. \emph{\textbf{(a)}} The logarithmic derivative of the one-electron WF.~\emph{\textbf{(b)}} The logarithmic estimate of the accuracy of the different one-electron WF approximations.}
\includegraphics[width=8.0in]{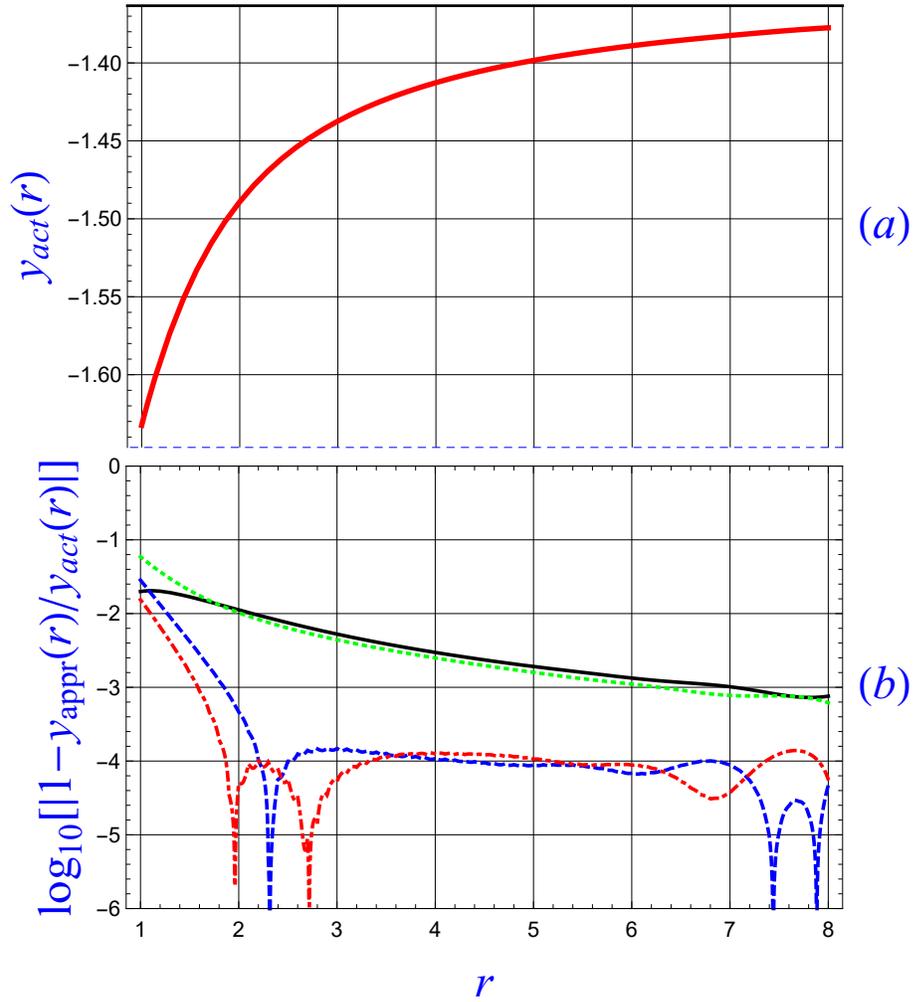}
\label{F2}
\end{figure}

In Fig.\ref{F2} we demonstrate the "actual" and the approximate one-electron WFs for the ground state of the helium atom.
Fig.\ref{F2}\emph{a} represents the logarithmic derivative $y_{act}(r)$ of the "actual" one-electron WF obtained by the AWFC. Fig.\ref{F2}\emph{b} presents the estimation functions (\ref{J8}) for four kinds of approximation functions.
In particular, functions $\mathfrak{L}_1(0)$ and  $\mathfrak{L}_1(\tau_1)$ characterizing the accuracy of the approximate WF (\ref{J4}) without shift and with the shift $\tau_1\simeq 0.2371$ are depicted by solid (black online) and dashed (blue online) curves, respectively.
Similarly, functions $\mathfrak{L}_2(0)$ and  $\mathfrak{L}_2(\tau_2)$ characterizing the accuracy of the approximate WF (\ref{J7}) without and with the shift $\tau_2\simeq - 0.1765$ are depicted by dotted (green online) and dot-dashed (red online) curves, respectively.

\begin{figure}
\centering
\caption{The ground state of the positive ion of boron $\text{B}^{3+}$. \emph{\textbf{(a)}} The logarithmic derivative of the one-electron WF.~\emph{\textbf{(b)}} The logarithmic estimate of the accuracy of the different one-electron WF approximations.}
\includegraphics[width=8.0in]{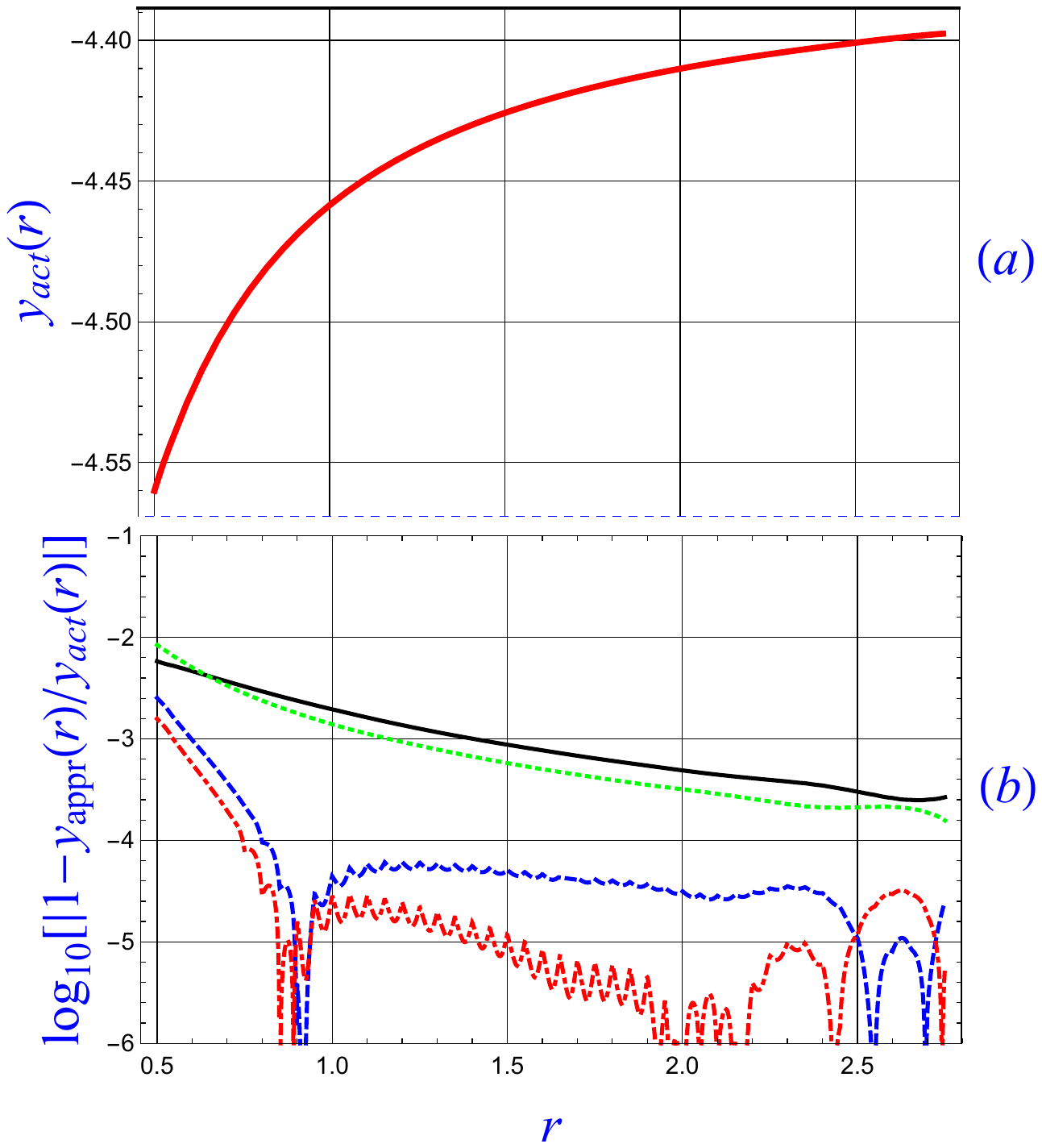}
\label{F3}
\end{figure}

The plots in Fig.\ref{F3} correspond to those in
Fig.\ref{F2}, but for the positive ion of boron $\text{B}^{3+}$ ($Z=5$) in the ground state.
The shifts of arguments for this case, as well as for all two-electron atoms with $1\leq Z\leq5$, are presented in Table \ref{T2}. All shifts are obtained by direct fitting the WFs calculated by the AWFC.
Note that the shifts associated with approximation (\ref{J7}) are negative.

\begin{figure}
\centering
\caption{Negative ion of hydrogen $\text{H}^{-}$. \emph{\textbf{(a)}} The logarithmic derivative of the one-electron WF.~\emph{\textbf{(b)}} The logarithmic estimate of the accuracy of the different one-electron WF approximations.}
\includegraphics[width=8.0in]{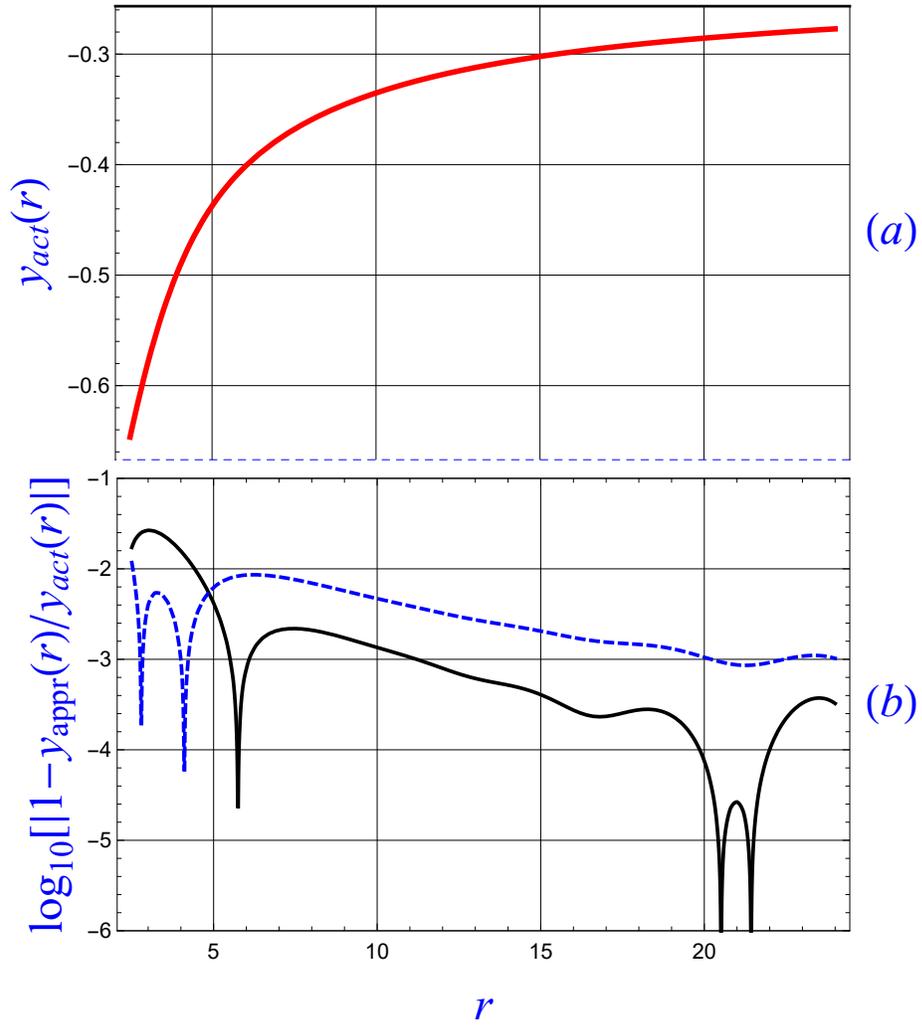}
\label{F4}
\end{figure}

In Fig.\ref{F4} we present the "actual" and the approximate one-electron WFs for the negative ion of hydrogen  $\text{H}^{-}$ ($Z=1$). It was mentioned earlier that the approximation (\ref{J7}) is not applicable to the two-electron ion with $Z=1$. Therefore, only the approximation of the form (\ref{J4}) with and without shift is depicted in this figure.

The figures present only the limiting cases of $Z$ due to the limited volume of the paper.
It is seen that the approximations with shifts are substantially more accurate than the ones without shifts for $Z>1$. Such increased accuracy enables one to apply approximations (\ref{J4}) and (\ref{J7}) with the shifts, presented in Table \ref{T2}, not only for very large values of $r$ but, e.g., for $r>1.5$ in case of $\text{He}$, or $r>0.5$ for the positive ion of boron $\text{B}^{3+}$. It is seen from Fig.\ref{F4} that in the case of the negative ion of hydrogen $\text{H}^-$, the approximation (\ref{J4}) with shift is more accurate for $r<5$, unlike the approximation without shift.

\subsection{Behavior near the nucleus} \label{S2b}

It was shown in Ref.\cite{HOF3} that the logarithmic derivative  of the one-electron density at the nucleus of the $n$-electron atomic system is defined as
\begin{equation}\label{J9}
\rho'(0)/\rho(0)=-2 Z,
\end{equation}
where $Z$ is the nuclear charge (see also \cite{KATO},\cite{HOF1} and \cite{FOUR1}).
It has been proven in Ref.\cite{HOF4} that $\rho''(0)$ exists and is non-negative.
In Ref.\cite{FOUR2} the existence of $\rho'''(0)$ has also been proven.

The AWFC shows that the relationship (\ref{J9}) is satisfied with high accuracy for the Pekeris-like calculations \cite{LEZ1} (with the number of shells $\Omega=25$).
The results of such calculations for the relative second derivative at the nucleus are presented in Table \ref{T3}.
However, we suspect that calculations of $\rho''(0)/\rho(0)$ by the Pekeris-like method may not be accurate enough.
Although this method can apply a large enough basis (1729 basis functions in the present case), this basis doesn't contain the logarithmic functions which are very important near the nucleus \cite{FOCK}.
Therefore, we have also presented in Table \ref{T3} the results of the same calculations performed by the correlation-function hyperspherical-harmonic  method (CFHHM) \cite{HMA} which
yields a very accurate numerical representation of the WFs near the nucleus, based on its correct analytical structure.
We used CFHHM with $K_m=128$ (basis size $N=1089$) for $Z=1$, and  $K_m=96$ ($N=625$) for $Z>1$.
The results of both methods are very close, but we tend to believe that the results (under consideration) obtained by the CFHHM are more reliable.

Let's consider the model WF of the form
\begin{equation}\label{J10}
\psi_1(\lambda)=\exp\left[-Z(r_1+r_2)+\lambda r_{12}\right],
\end{equation}
which for $\lambda=1/2$ was studied in Ref.\cite{MYERS} as satisfying the two-particle Kato's cusp conditions.
Note that the relation $r_{12}=\sqrt{r_1^2+r_2^2-2 r_1 r_2 \omega}$ enables us to express $\psi_1$ in the $\left\{r_1,r_2,\omega\right\}$ coordinate system. Putting $\Phi=\psi_1(1/2)$ in the definition (\ref{J3}), we obtain
\begin{equation}\label{J11}
\rho''(0)/\rho(0)=\frac{2}{3}Z(6Z+1).
\end{equation}
Numerical results corresponding to Eq.(\ref{J11}) are presented in Table \ref{T3}. It is seen that these results are close but slightly larger then the "actual" ones. On the other hand, for the simplest model WF of the form $\psi_1(0)$ we obtain
\begin{equation}\label{J12}
\rho''(0)/\rho(0)=4Z^2.
\end{equation}
 The corresponding numerical results presented in Table \ref{T3} show that formula (\ref{J12}) yields $\rho''(0)/\rho(0)$ which are close to, but slightly smaller than the "actual" ones.
Thus, we can conclude that the model WFs mentioned above produce the "correct" values for the upper and lower bounds of the "actual" values of $\rho''(0)/\rho(0)$.
Note that both $\psi_1(1/2)$ and $\psi_1(0)$ produce logarithmic derivatives which satisfy Eq.(\ref{J9}).

The model WF $\psi_1(\lambda)$ with presently undetermined parameter $\lambda < Z$ generates (by Eq.(\ref{J3})) the relative second derivative (at the nucleus) of the form:
\begin{equation}\label{J13}
\rho_1''(0)/\rho_1(0)=\frac{4}{3}Z(\lambda+3Z).
\end{equation}
It is natural to suppose that a suitable choice of the parameter $\lambda$ can produce
$\rho_1''(0)/\rho_1(0)$ close to the actual value.
The method of finding  $\lambda$ is the following. In Appendix \ref{SB} we have derived the differential equation (\ref{B3}) for the one-electron density $\rho(r)$ of the two-electron atom/ions. Substitution of the model density $\rho(r)\equiv\rho_1(r)$ into Eq.(\ref{B3}) and the subsequent use of the power series expansions of all the terms in the vicinity of the nucleus ($r=0$) yields the following results. Collecting all the coefficients for $r^{-1}$, we obtain Eq.(\ref{J9}). Collecting all the coefficients for $r^{0}$ (and using Eq.(\ref{J9}) ), we obtain the relationship
\begin{equation}\label{J14}
\lambda=\frac{1}{2}\left(Z+1-\sqrt{5Z^2-2Z+4E+1}\right).
\end{equation}
The numerical results corresponding to Eqs.(\ref{J13}) and (\ref{J14}) are shown in Table \ref{T3}.
It is seen that the corresponding values of $\rho_1''(0)/\rho_1(0)$ are rather close to but exceed the actual by $\sim 0.2$.

It is seen that the model WF (\ref{J10}) is symmetric regarding permutation of the electrons $(r_1\leftrightarrows r_2)$.
This property is determinative for the ground state of the atom/ion.
The more general model WF possessing the same symmetry properties is of the form (to within a constant multiplicative factor):
\begin{equation}\label{J15}
\psi_2(a,b,\lambda)=\exp(\lambda r_{12})\left[\exp(-a r_1-b r_2)+\exp(-a r_2-b r_1)\right].
\end{equation}
It is worth noting that by putting $\Phi=\psi_2$ in the definition (\ref{J3}),  we can derive the simple representation for the  one-electron density in the explicit form:
\begin{eqnarray}\label{J16}
\rho_2(r)=
\frac{\lambda}{2}e^{-2r(a+b)}\left\{\frac{a \left(a+\frac{2}{r}\right)-\lambda^2}{(\lambda^2-a^2)^3}
+\frac{32\left[4 \lambda^2-(a+b)(a+b+\frac{4}{r})\right]}{\left[(a+b)^2-4\lambda^2\right]^3}
+\frac{b \left(b+\frac{2}{r}\right)-\lambda^2}{(\lambda^2-b^2)^3}
\right\}+~~\nonumber~~~\\
e^{-2r(b-\lambda)}\frac{a\left(\lambda^2-a^2-\frac{2\lambda}{r}\right)}{2(\lambda^2-a^2)^3}+
e^{-2r(a-\lambda)}\frac{b\left(\lambda^2-b^2-\frac{2\lambda}{r}\right)}{2(\lambda^2-b^2)^3}+
~~~~~~~\nonumber~~~~~~~~~~~~~~~~~~~~~~~~~\\
e^{-r(a+b-2\lambda)}\left\{1+\frac{8\lambda}{r\left[(a+b)^2-4\lambda^2\right]}\right\}\frac{8(a+b)}{\left[(a+b)^2-4\lambda^2\right]^2}.
~~~~~~~~~
\end{eqnarray}
Using Eq.(\ref{J16}) we easily obtain the expressions for the model density $\rho_2$ and its derivatives at the nuclei:
\begin{equation}\label{J17}
\rho_2(0)=\frac{1}{2}\left[\frac{16}{(a+b-2\lambda)^3}+\frac{1}{(a-\lambda)^3}+\frac{1}{(b-\lambda)^3}\right],
\end{equation}
\begin{equation}\label{J18}
\rho_2'(0)=-\frac{8(a+b)}{(a+b-2\lambda)^3}-\frac{b}{(a-\lambda)^3}-\frac{a}{(b-\lambda)^3},
\end{equation}
\begin{equation}\label{J19}
\rho_2''(0)=\frac{2}{3}\left\{\frac{4(a+b)[3(a+b)+2\lambda]}{(a+b-2\lambda)^3}+
\frac{3b^2+a\lambda}{(a-\lambda)^3}+\frac{3a^2+b \lambda}{(b-\lambda)^3}\right\}.
\end{equation}
Formulas for the corresponding relative derivatives follow directly from Eqs.(\ref{J17})-(\ref{J19}).

The appropriate choice of the parameters $a,b,\lambda$ in the model WF (\ref{J15}) enables us to obtain the model one-electron density (\ref{J16}) representing a very accurate approximation for the "actual" densities of the helium-like atomic systems.
In particular, it was found that one of the symmetric parameters ($a$ and $b$), e.g., $a$ can be taken in the form $a=Z-1/2$.
Simulation of the "actual" densities by the model density (\ref{J16}) enables us to calculate two other parameters $b$ and $\lambda$ presented in Table \ref{T4} together with the corresponding relative derivatives at the nuclei.
It is seen that these derivatives are very close to but not accurate enough in comparison with the "actual" ones.

There is another way for deriving the form of the model density function $\rho_3(r)$ giving a very good approximation (not for large $r$) to the "actual" one including the relative derivatives at the nuclei.

Let's suppose that Eq.(\ref{B3}) for the one-electron density $\rho(r)$ can be approximately replaced with the general equation
\begin{equation}\label{J20}
\rho_3''(r)+\left(\frac{A}{r}+B\right)\rho_3'(r)+\left(\frac{C}{r}+D\right)\rho_3(r)=0
\end{equation}
for the model density $\rho_3(r)\simeq \rho(r)$, where four parameters $A,B,C,D$ are presently undetermined.
Remember that in Sec.\ref{S1} we treated a similar equation but for the WF $\Psi$ averaged over the hyperspherical angles.
Given that, as was mentioned earlier, at least the first three derivatives $\rho'(0),\rho''(0)$ and $\rho'''(0)$ exist,
we can use the power series expansion of all terms of Eq.(\ref{J20}) near the nucleus ($r=0$).
Thus, equating the coefficients for $r^{-1}$ on both sides of Eq.(\ref{J20}) one obtains the relation $A\rho_3'(0)+C\rho_3(0)=0$, whence using Eq.(\ref{J9}) we obtain the first coupling equation
\begin{equation}\label{J21}
C=2 Z A
\end{equation}
for the coefficients contained in this equation. Collecting the coefficients for $r^0$ in Eq.(\ref{J20}) and using Eqs.(\ref{J9}) and (\ref{J21}), we obtain the second coupling equation
\begin{equation}\label{J22}
(1+A)\frac{\rho_3''(0)}{\rho_3(0)}-2Z(B+2 Z A)+D=0.
\end{equation}
The general solution of Eq.(\ref{J20}) is of the form
\begin{equation}\label{J24}
\rho_3(r)= e^{-\frac{1}{2}r(B+\sigma)}\left[c_1 U(\kappa,A,\sigma r)+c_2 L_{-\kappa}^{A-1}(\sigma r)\right],
\end{equation}
where
\begin{equation}\label{J25}
\sigma=\sqrt{B^2-4D},~~~~~~~~~~~\kappa=\frac{A}{2}\left(1+\frac{B-4Z}{\sigma}\right),
\end{equation}
and $U(a,b,z)$ and $L_{-a}^{b-1}(z)$ are again the Tricomi function and the generalized Laguerre function, respectively.
The general solution (\ref{J24}) therefore depends on three parameters only: $A,B$ and $D$.
Considering the relative second derivative $\rho_3''(0)/\rho_3(0)$ in the coupling equation (\ref{J22})
 as a known constant, the number of free parameters reduces to two.

It is shown in Appendix \ref{SC} that we should set $c_1=0$ in the general solution (\ref{J24}) to obtain the physical solution (to within a constant multiplicative factor) of the form:
\begin{equation}\label{J26}
\rho_3(r)= e^{-\frac{1}{2}r(B+\sigma)} L_{-\kappa}^{A-1}(\sigma r).
\end{equation}
Parameters $A$ and $B$ obtained by fitting the "actual" one-electron densities are presented in Table \ref{T4}. Parameters $C$ and $D$
can be calculated from Eqs.(\ref{J21}) and (\ref{J22}), respectively. For these calculations we used the "actual" relative second derivatives (at the nuclei) obtained with the use of WFs calculated by the CFHHM \cite{HMA}, instead of $\rho_3''(0)/\rho_3(0)$ (see Eq.(\ref{J22})).
It is easy to verify that all $D$-coefficients, calculated by Eq.(\ref{J22}) with the use of the coefficients $A$ and $B$ presented in Table \ref{T4}, are positive. It is shown in Appendix \ref{SC} that in this case the model solution (\ref{J26}) tends exponentially to zero as $r\rightarrow \infty$, however, the corresponding asymptotic behavior is far away from accurate.
Note that finding the one-electron densities with very accurate asymptotic behavior was discussed
in Sec.\ref{S2a}.

\begin{figure}
\centering
\caption{Negative ion $\text{H}^{-}$. \emph{\textbf{(a)}} One-electron density $\rho(r)$ multiplied by $r$. \emph{\textbf{(b)}} The logarithmic estimate of the accuracy of the one-electron density approximations $\rho_2(r)$ and $\rho_3(r)$ depicted by dashed (blue online) and solid (red online) curves, respectively.}
\includegraphics[width=8.0in]{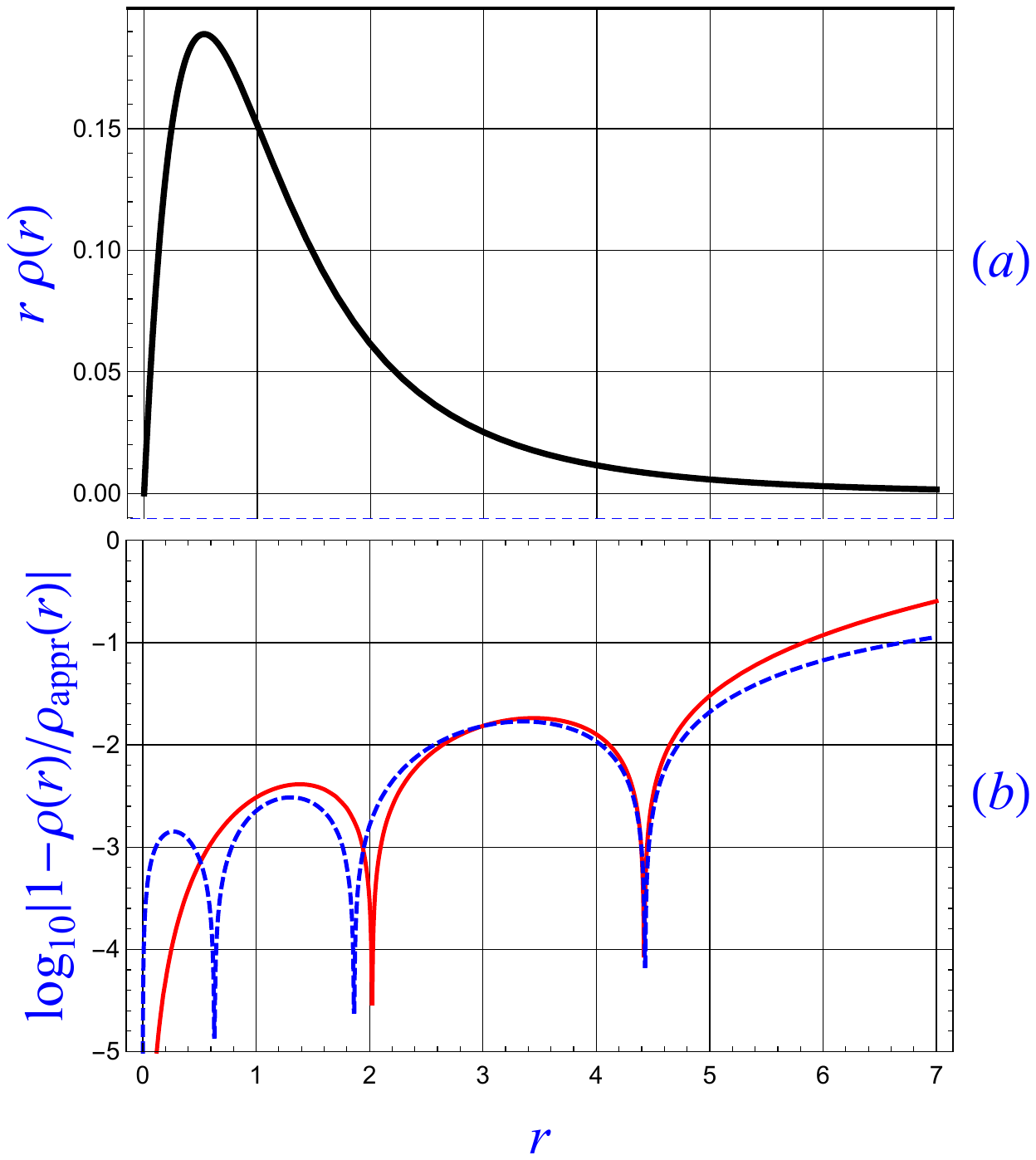}
\label{F5}
\end{figure}

\begin{figure}
\centering
\caption{Helium atom. \emph{\textbf{(a)}} One-electron density $\rho(r)$ multiplied by $r$. \emph{\textbf{(b)}} The logarithmic estimate of the accuracy of the one-electron density approximations $\rho_2(r)$ and $\rho_3(r)$ depicted by dashed (blue online) and solid (red online) curves, respectively.}
\includegraphics[width=8.0in]{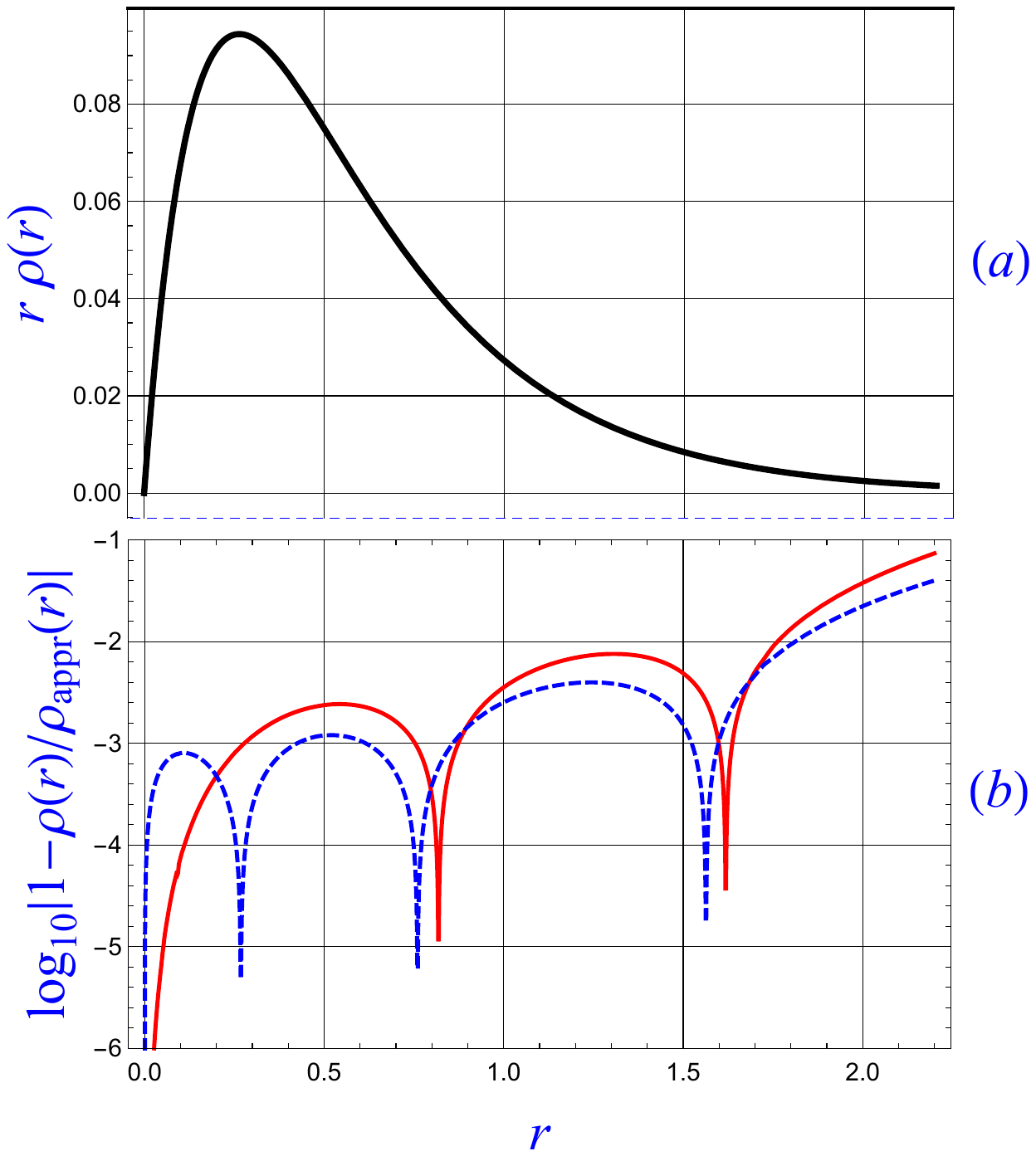}
\label{F6}
\end{figure}

\begin{figure}
\centering
\caption{Positive ion $B^{3+}$. \emph{\textbf{(a)}} One-electron density $\rho(r)$ multiplied by $r$. \emph{\textbf{(b)}} The logarithmic estimate of the accuracy of the one-electron density approximations $\rho_2(r)$ and $\rho_3(r)$ depicted by dashed (blue online) and solid (red online) curves, respectively.}
\includegraphics[width=8.0in]{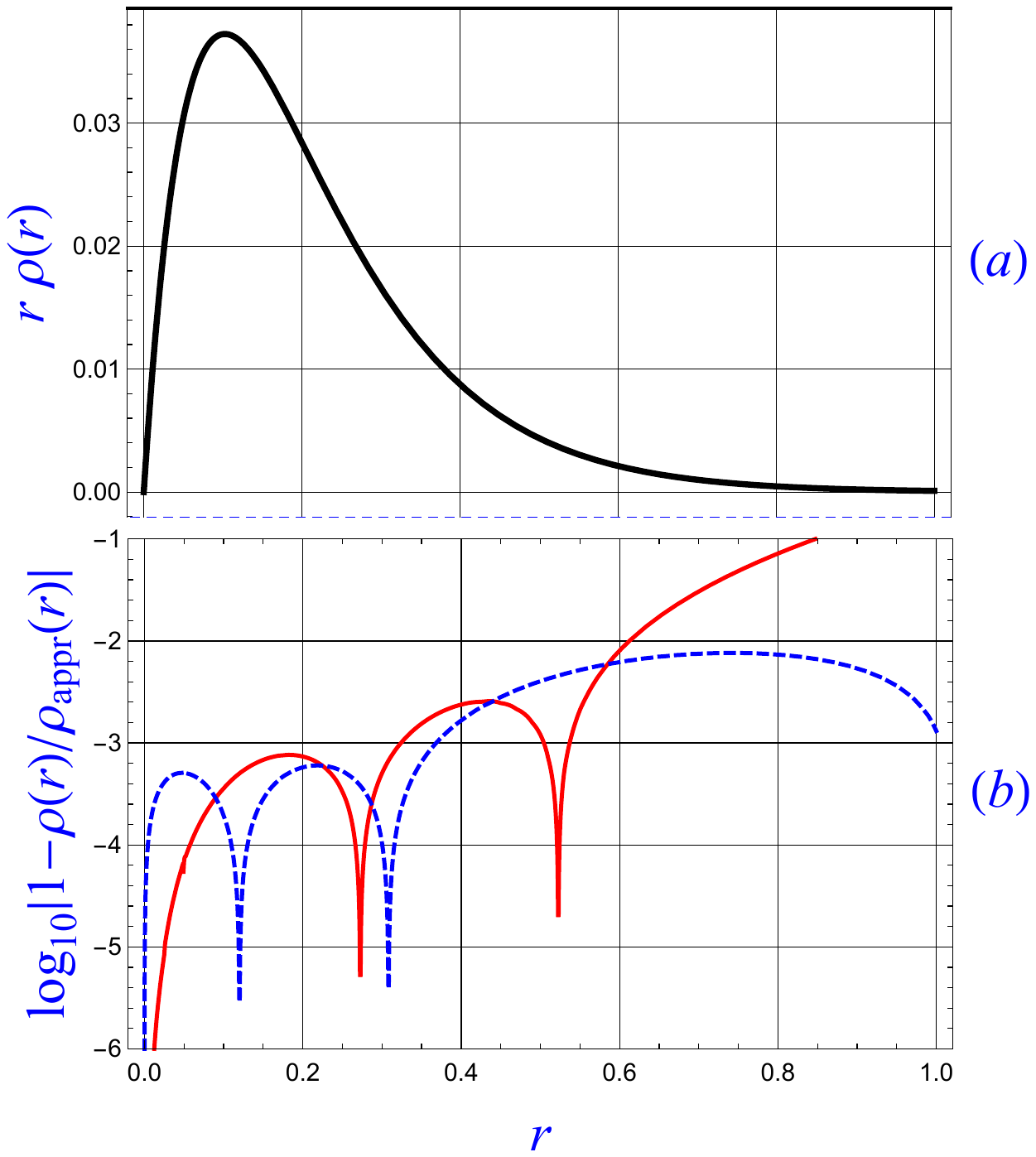}
\label{F7}
\end{figure}

In figures \ref{F5}-\ref{F7} we present the "actual" one-electron densities and their approximations represented by the
model densities $\rho_2(r)$ and $\rho_3(r)$ for the most characteristic helium-like atoms, namely for the negative ion $H^-$, atom of $He$ and positive ion $B^{3+}$.
Plots  $(a)$ on these figures represent the "actual" one-electron densities with $\rho(0)=1$.
Plots $(b)$ represent the accuracy of the model densities  $\rho_2(r)$ depicted by dashed (blue online) curve, and $\rho_3(r)$ depicted by solid (red online) curve. The accuracies mentioned above are described by the logarithmic functions $\mathfrak{L}_k(0)$ ($k=2,3$, see Eq.(\ref{J8})) with the  replacement of $y_{appr}$ and $y_{act}$ by $\rho_k$ and $\rho_{act}$, respectively. It is seen that for the negative ion of hydrogen the model density $\rho_3(r)$ is more accurate than  $\rho_2(r)$, whereas for $Z>1$ the latter density is more preferable. Moreover, figures (\ref{F5}$b$),(\ref{F6}$b$) and(\ref{F7}$b$) demonstrate a high accuracy of both model functions not only for small electron-nucleus distance $r$ but in its intermediate range too. Thus, matching (at some intermediate point) these model densities with the model ones described in Sec.\ref{S2a} enables us to obtain very accurate analytic approximations for the one-electron densities of the two-electron atomic systems.

\section{Conclusions} \label{S3}

The use of the Pekeris-like method \cite{LEZ1} (with number of shells $\Omega=25$) enables us to calculate
the ground state wave functions $\Psi$ of the helium-like atom/ions with nucleus charge $1\leq Z\leq5$.
The standard coordinate system $\left\{r_1,r_2,r_{12}\right\}$ of the interparticle distances, as well as the hyperspherical coordinates $\left\{R,\alpha,\theta\right\}$ and the coordinates $\left\{r_1,r_2,\cos \theta \right\}$
were applied.

The averaged wave functions $\Psi_{av}(R)$ of the hyperspherical radius $R$ were calculated numerically by averaging $\Psi(R,\alpha,\theta)$ over the hyperspherical angles $\alpha$ and $\theta$.
The exact analytic representations for the relative derivatives $\Psi_{av}'(0)/\Psi_{av}(0)$ and $\Psi_{av}''(0)/\Psi_{av}(0)$ were derived by the use of the angular coefficients \cite{LEZ2} of the Fock expansion \cite{FOCK}.
The analytic approximations very close to the actual $\Psi_{av}(R)$ were obtained on the base of the specific form of the Schr\"{o}dinger equation for the two-electron atoms.

Using the actual wave functions $\Psi(r_1,r_2,\cos \theta)$, the one-electron densities $\rho(r_1)\equiv \rho(r_2)$ were calculated.
The first and second relative derivatives $\rho'(0)/\rho(0)$ and $\rho''(0)/\rho(0)$ at the nuclei were calculated numerically by the use of "actual" WFs which have been obtained by the Pekeris-like method \cite{LEZ1} as well as the CFHHM \cite{HMA}.
Very accurate analytical approximations, representing the actual one-electron density both near the nucleus and far away from it, were derived. It is worth noting that both approximations are accurate enough in the intermediate region too which enables us to match them at some point $r=r_0$ inside this region
to obtain the analytic approximation describing the one-electron density $\rho(r)$
in the full range $r\in [0,\infty]$.

All the analytical and numerical results are illustrated by tables \ref{T1}-\ref{T4} and graphs \ref{F1}-\ref{F7}.

\section{Acknowledgment}

This work was supported by the PAZY Foundation.

\appendix

\section{}\label{SA}

The final representations (\ref{K10}) and (\ref{K11}) for the first and second relative derivatives of the averaged WF at $R=0$ can be certainly calculated by direct integration of the angular coefficients $\psi_{1,0}(\alpha,\theta)$ and  $\psi_{2,0}(\alpha,\theta)$ according to definition (\ref{K4}). The other way consists of applying the RER (\ref{K8}).
For the first derivative it is more convenient to use the direct approach.
First of all, note that the AC $\psi_{0,0}$ is a constant, the exact value of which does not matter for our consideration.
According to (\ref{K9}), the AC required for calculation of $\Psi_{av}'(0)$ is (see, e.g., \cite{LEZ2}):
\begin{equation}\label{A1}
\psi_{1,0}=-Z\sqrt{1+\sin \alpha}+\frac{1}{2}\sqrt{1-\sin \alpha \cos \theta}.
\end{equation}
Whence we obtain the relation
\begin{equation}\label{A2}
(\psi_{1,0})_{av}=\int_0^{\pi}d\theta\int_0^{\pi}\psi_{1,0}(\alpha,\theta)\sin^2\alpha \sin \theta~d\alpha=-\frac{16(4Z-\sqrt{2})}{15}.
\end{equation}
Taking into account the extra relation
\begin{equation}\label{A3}
(\psi_{0,0})_{av}=\int_0^{\pi}d\theta\int_0^{\pi}\psi_{0,0}(\alpha,\theta)\sin^2\alpha \sin \theta~d\alpha=\pi \psi_{0,0}
\end{equation}
we obtain the final result (\ref{K10}).

For deriving the second derivative $\Psi_{av}''(0)$, application of the RER (\ref{K8}) is more preferable (simple).
First of all, the equality $(\psi_{2,1})_{av}=0$ follows directly from the recurrence relations (\ref{K8}) for $k=2\wedge p=1$. Integration of the AC  $\psi_{2,1}(\alpha,\theta)$ the explicit form of which can be found, e.g., in Ref.\cite{LEZ2}, yields certainly the same result.
For  $k=2\wedge p=0$ the RER (\ref{K8}) becomes:
\begin{equation}\label{A4}
12(\psi_{2,0})_{av}-2(\mathcal{V}\psi_{1,0})_{av}+2 E (\psi_{0,0})_{av}=0.
\end{equation}
It follows from definition (\ref{K3}) that the potential $\mathcal{V}=R~V$ is of the form
\begin{equation}\label{A5}
\mathcal{V}=-\frac{2Z\sqrt{1+\sin \alpha}}{\sin \alpha}+\frac{1}{\sqrt{1-\sin \alpha \cos \theta}}
\end{equation}
in the hyperspherical coordinates (\ref{I1}). Using the explicit forms (\ref{A1}) and (\ref{A5}) we obtain:
\begin{equation}\label{A6}
\left(\mathcal{V}\psi_{1,0}\right)_{av}=\int_0^{\pi}d\theta\int_0^{\pi}\mathcal{V}(\alpha,\theta)\psi_{1,0}(\alpha,\theta)\sin^2\alpha \sin \theta~d\alpha=\frac{1}{2}(2Z-1)\left[8Z+\pi(2Z-1)\right].~~
\end{equation}
Substitution of Eqs.(\ref{A6}) and (\ref{A3}) into the RER (\ref{A4}) finally yields Eq.(\ref{K11}).

\section{}\label{SB}

Our aim is to find an analytic expression for the one-electron density $\rho(r)$ that can be quite accurate not only for very large values of $r$ but in the intermediate range too.

The  Schr\"{o}dinger equation for the two-electron atoms can be presented in the form
\begin{equation}\label{B1}
\Delta\Phi+2 (E-V)\Phi=0,
\end{equation}
where in the $\left\{r_1,r_2,\omega\right\}$ coordinate system the Laplacian $\Delta$ is of the form (\ref{J2}), and the Coulomb inter-particle interaction is:
\begin{equation}\label{B2}
V=-\frac{Z}{r_1}-\frac{Z}{r_2}+\frac{1}{\sqrt{r_1^2+r_2^2-2 r_1 r_2 \omega}}.
\end{equation}
Left multiplying Eq.(\ref{B1}) by $r_2^2\Phi^*$ with subsequent integration over $r_2$ and $\omega$ yields the equation
\begin{equation}\label{B3}
I_{1a}(r)+\frac{1}{r}I_{1b}(r)+\frac{1}{r^2}I_{1c}(r)+I_{2a}(r)+2I_{2b}(r)+I_{2c}(r)+2 Z v_2(r)+2 v_3(r)=
-2\left(\frac{Z}{r}+E\right)\rho(r),
\end{equation}
where $\rho$ is defined by Eq.(\ref{J3}), and the other terms of Eq.(\ref{B3}) are:
\begin{equation}\label{B4}
I_{1a}(r)=\int_{-1}^1 d\omega\int_0^\infty \Phi^*(r,r_2,\omega)\frac{\partial^2 \Phi(r,r_2,\omega)}{\partial r^2} r_2^2 dr_2
=\frac{1}{2}\rho''(r)-J(r),
\end{equation}
\begin{equation}\label{B5}
J(r)=\int_{-1}^1 d\omega\int_0^\infty \left|\frac{\partial \Phi(r,r_2,\omega)}{\partial r}\right |^2 r_2^2 dr_2,
\end{equation}
\begin{equation}\label{B6}
I_{1b}(r)=2\int_{-1}^1 d\omega\int_0^\infty \Phi^*(r,r_2,\omega)\frac{\partial \Phi(r,r_2,\omega)}{\partial r} r_2^2 dr_2=\rho'(r),
\end{equation}
\begin{equation}\label{B7}
I_{2a}(r)=\int_{-1}^1 d\omega\int_0^\infty \Phi^*(r,r_2,\omega)\frac{\partial^2 \Phi(r,r_2,\omega)}{\partial r_2^2} r_2^2 dr_2,
\end{equation}
\begin{equation}\label{B8}
I_{2b}(r)=\int_{-1}^1 d\omega\int_0^\infty \Phi^*(r,r_2,\omega)\frac{\partial \Phi(r,r_2,\omega)}{\partial r_2} r_2 dr_2,
\end{equation}
\begin{equation}\label{B9}
I_{1c}(r)=\int_{-1}^1 d\omega\int_0^\infty \Phi^*(r,r_2,\omega)
\left[(1-\omega^2)\frac{\partial^2 \Phi(r,r_2,\omega)}{\partial\omega^2}-2\omega\frac{\partial \Phi(r,r_2,\omega)}{\partial\omega}\right] r_2^2 dr_2,
\end{equation}
\begin{equation}\label{B10}
I_{2c}(r)=\int_{-1}^1 d\omega\int_0^\infty \Phi^*(r,r_2,\omega)
\left[(1-\omega^2)\frac{\partial^2 \Phi(r,r_2,\omega)}{\partial\omega^2}-2\omega\frac{\partial \Phi(r,r_2,\omega)}{\partial\omega}\right] dr_2,
\end{equation}
\begin{equation}\label{B11}
v_2(r)=\int_{-1}^1 d\omega\int_0^\infty\left| \Phi(r,r_2,\omega)\right|^2 r_2 dr_2,
\end{equation}
\begin{equation}\label{B12}
v_3(r)=-\int_{-1}^1 d\omega\int_0^\infty\left| \Phi(r,r_2,\omega)\right|^2\frac{r_2^2}{\sqrt{r^2+r_2^2-2 r r_2 \omega}} dr_2.
\end{equation}
It was mentioned in Sec.\ref{S2a} that
we determine the behavior of $\rho(r)$ at different specific values of $r$ by the use of the "actual" WFs \cite{LEZ1} for numerical calculations (AWFC) of the one-electron densities/wavefunctions.
Thus, first of all, the AWFC shows that $I_{1a}(r)-J(r)$ approaches zero as $r\rightarrow \infty$, which implies
that
\begin{equation}\label{B13}
I_{1a}(r) =\frac{1}{4}\rho''(r).
\end{equation}
at large enough $r$. Second, the AWFC shows the following properties of the integrals under consideration:
\begin{equation}\label{B14}
I_{2a}(\infty)/\rho(\infty) =Z^2,~~~~
I_{2b}(\infty)/\rho(\infty) =-Z^2,~~~~
v_{2}(\infty)/\rho(\infty) =Z.
\end{equation}
Given the asymptotic ($r\rightarrow \infty$) expansion of the electron-electron interaction
\begin{equation}\label{B15}
\frac{1}{\sqrt{r^2+r_2^2-2 r r_2 \omega}}\underset{r\rightarrow \infty}{=}\frac{1}{r}+\frac{r_2 \omega}{r^2}+O\left(r^{-3}\right),
\end{equation}
and neglecting the terms of the order $r^{-m}$ with $m\geq2$ for large enough $r$, we obtain
\begin{equation}\label{B16}
v_3(r)\underset{r\rightarrow \infty}{=}-\frac{1}{r}\rho(r).
\end{equation}
And at last, the AWFC shows that we can neglect the terms $I_{1c}(r)/r^2$ and $I_{2c}(r)$ for large enough electron-nucleus distance $r$. Applying the properties described above we obtain the equation (\ref{J5}).

\section{}\label{SC}

The power series expansions for the Tricomi function appearing in the solution (\ref{J24}) are:
\begin{equation}\label{C1}
U(\kappa,A,\sigma r)\underset{r\rightarrow 0}{=}(r \sigma)^{-A}
\left[\frac{\sigma\Gamma(A-1)r}{\Gamma(\kappa)}+O(r^2)\right]+
\frac{\Gamma(1-A)}{\Gamma(1-A+\kappa)}\left[1+\frac{\kappa \sigma r}{A}+O(r^2)\right],
\end{equation}
\begin{equation}\label{C2}
U(\kappa,A,\sigma r)\underset{r\rightarrow \infty}{=}(r \sigma)^{-\kappa}
\left[1-\frac{\kappa(1-A+\kappa)}{\sigma r}+O\left(\frac{1}{r^2}\right)\right].~~~~~~~~~~~~~~~~~~~
\end{equation}
It is seen that the asymptotic expansion (\ref{C2}) can provide the physical behavior of the density $\rho_3(r)$ as $r\rightarrow\infty$ for any set of parameters (see the general solution (\ref{J24})), whereas the near-the-origin expansion (\ref{C1}) cannot provide the physical behavior of the density $\rho_3(r)$ as $r\rightarrow 0$ ($\rho(0)=const$) for any set of parameters.

The power series expansions for the generalized Laguerre function included into the solution (\ref{J24}) are:
\begin{equation}\label{C3}
L_{-\kappa}^{A-1}(\sigma r)\underset{r\rightarrow 0}{=}\frac{\Gamma(A-\kappa)}{\Gamma(A+1)\Gamma(1-\kappa)}\left(A+\sigma \kappa r+O(r^2)\right),~~~~~~~~~~~~~
\end{equation}
\begin{eqnarray}\label{C4}
L_{-\kappa}^{A-1}(\sigma r)\underset{r\rightarrow \infty}{=}
\frac{(-\sigma r)^{-\kappa}}{\Gamma(1-\kappa)}
\left[1+\frac{\kappa(A-\kappa-1)}{\sigma r}+O\left(\frac{1}{r^2}\right)\right]+
~~~~~~~~~\nonumber~~~~~~~~~~~~~~~~~~~~~~~~~\\
e^{\sigma r}(\sigma r)^{\kappa-A}\frac{\Gamma(A-\kappa)\sin(\pi \kappa)}{\pi}\left[1+\frac{(\kappa-1)(\kappa-A)}{\sigma r}+O\left(\frac{1}{r^2}\right)\right].~~~~~~~~~~~
\end{eqnarray}
It is seen that the near-the-origin expansion (\ref{C3}) can provide the physical behavior of the density $\rho_3(r)$ as $r\rightarrow 0$ for any set of parameters, whereas the asymptotic expansion (\ref{C4}) can provide the physical behavior of the density $\rho_3(r)$ as $r\rightarrow \infty$ only under the condition
\begin{equation}\label{C5}
B>\sigma.
\end{equation}
The case of $A-\kappa$ being equal to a non-positive integer is certainly excluded too.

It follows from the definition of $\sigma$ in Eq.(\ref{J25}) that the inequality (\ref{C5}) can be satisfied only under the condition
\begin{equation}\label{C6}
D>0.
\end{equation}

\newpage

\begin{table}
\caption{Parameters of the approximation (\ref{K16}), (\ref{K17}) of the averaged WFs for the ground state of the two-electron atom/ions with nucleus charge $Z$.}
\begin{tabular}{|c|c|c|c|c|c|}
\hline
$Z$ & $A_{0}$ & $A_{1}$ & $B_{0}$ & $B_{1}$ & $R_{m}$\tabularnewline
\hline
\hline
1 & 0.870771  & -1.70712 &  1.87387  & -1.94443 & 10\tabularnewline
\hline
2 & 5.42995 &  -4.81369  & 4.65702 &  -2.15274  & 5\tabularnewline
\hline
3 & 13.3998 & -7.81131 & 7.30891 &  -2.17331 & 2.5\tabularnewline
\hline
4 & 24.9767 & -10.5068 & 9.9784 & -2.12159 & 2\tabularnewline
\hline
5 & 38.8192  &  -10.5392  & 12.4492 &  -1.67012 & 1.5\tabularnewline
\hline
\end{tabular}
\label{T1}
\end{table}

\begin{table}
\caption{The shifts $\tau_1$ and $\tau_2$ of arguments of the approximate WFs (\ref{J4}) and (\ref{J7}), respectively.}
\begin{tabular}{|c|c|c|c|c|c|}
\hline
 & $\text{H}^-$ & $\text{He}$ & $\text{Li}^+$ & $\text{Be}^{2+}$ & $\text{B}^{3+}$\tabularnewline
\hline
\hline
$\tau_1$ & 0.1112  & 0.2371 & 0.1434  & 0.1110 & 0.0922\tabularnewline
\hline
$\tau_2$ & $-$  & -0.1765  & -0.1123 &  -0.07709  & -0.0587\tabularnewline
\hline
\end{tabular}
\label{T2}
\end{table}

\begin{table}
\caption{The relative second derivative $\rho''(0)/\rho(0)$ of the one-electron WF at the nucleus of the charge $Z$.}
\begin{tabular}{|c|c|c|c|c|c|}
\hline
 $WF\setminus Z$& $1$ & $2$ & $3$ & $4$ & $5$\tabularnewline
\hline
\hline
 Actual (Pekeris-like \cite{LEZ1})& 4.173  & 16.707 & 37.214  & 65.716 & 102.218 \tabularnewline
\hline
 Actual (Haftel-Mandelzweig \cite{HMA})& 4.175  & 16.711 & 37.219  & 65.724 & 102.23 \tabularnewline
\hline
 $\psi_1(1/2)$ & 4.667  & 17.333 & 38 & 66.667 & 103.333 \tabularnewline
\hline
$\psi_1(0)$ & 4  & 16 & 36 & 64 & 100 \tabularnewline
\hline
$\psi_1$ & 4.417  & 16.906 & 37.403 & 65.901 & 102.401 \tabularnewline
 $(\lambda)$& (0.312796)  & (0.339709)& (0.350731) &(0.356537) & (0.360108) \tabularnewline
\hline
\end{tabular}
\label{T3}
\end{table}

\begin{table}
\caption{The relative derivatives at the nuclei
and parameters of the model one-electron densities $\rho_2(r)$ and $\rho_3(r)$ defined by Eq.(\ref{J16}) and Eq.(\ref{J26}), respectively.}
\begin{tabular}{|c|c|c|c|c|c|}
\hline
& $\text{H}^-$ & $\text{He}$ & $\text{Li}^+$ & $\text{Be}^{2+}$ & $\text{B}^{3+}$\tabularnewline
\hline
\hline
 $\rho_2'(0)/\rho_2(0)$& -2.012  & -4.017 & -6.022  & -8.024 &-10.025 \tabularnewline
\hline
$\rho_2''(0)/\rho_2(0)$&4.277   & 17.060 & 37.886  & 66.680 & 103.46 \tabularnewline
\hline
 $b$ &1.08908  & 2.24823 &3.32322 & 4.36538 & 5.39292 \tabularnewline
\hline
$\lambda$ & 0.17625  & 0.289378 & 0.343728 & 0.375608 & 0.395233 \tabularnewline
\hline
 $\rho_3'(0)/\rho_3(0)$& -2  & -4 & -6  & -8 &-10 \tabularnewline
\hline
$\rho_3''(0)/\rho_3(0)$&4.175   & 16.711 & 37.219  & 65.724 & 102.23 \tabularnewline
\hline
 $A$& 0.658135 & -1.33742 & -1.87079  & -2.12159 &-2.28372 \tabularnewline
\hline
$B$&3.03553   & 7.63567 & 11.9018  & 16.1157 & 20.3376\tabularnewline
\hline
\end{tabular}
\label{T4}
\end{table}

\end{document}